\documentclass[aps,prb,twocolumn,floatfix,showpacs]{revtex4}

\usepackage{graphicx}
\usepackage{bbm}
\usepackage{amsmath,amssymb}

\newcommand{\be}{\begin{equation}}
\newcommand{\beq}{\begin{equation}}
\newcommand{\ee}{\end{equation}}
\newcommand{\bea}{\begin{eqnarray}}
\newcommand{\eea}{\end{eqnarray}}
\newcommand{\ba}{\begin{array}}
\newcommand{\ea}{\end{array}}

\renewcommand{\vr} {{\bf r}}

\newcommand{\vk} {{\bf k}}

\begin{document}
\title{Constraints of reduced density-matrix functional theory
for the two-dimensional homogeneous electron gas}
\author{A. Putaja}
\affiliation{Nanoscience Center, Department of Physics, University of
  Jyv\"askyl\"a, FI-40014 Jyv\"askyl\"a, Finland}
\author{E. R{\"a}s{\"a}nen}
\email[Electronic address:\;]{erasanen@jyu.fi}
\affiliation{Nanoscience Center, Department of Physics, University of
  Jyv\"askyl\"a, FI-40014 Jyv\"askyl\"a, Finland}

\date{\today}

\begin{abstract}
Reduced density-matrix functional theory (RDMFT) has become an appealing
alternative to density-functional theory to describe 
electronic properties of highly-correlated systems. 
Here we derive exact conditions for the suitability of RDMFT to describe 
the two-dimensional homogeneous electron gas, which is the base 
system for, e.g., semiconductor quantum dots and quantum Hall devices.
Following the method of Cioslowski and Pernal 
[J. Chem. Phys. {\bf 111}, 3396 (1999)] we focus on the properties
of power functionals of the form $f(n,n')=(n n')^\alpha$ for the
scaling function in the exchange-correlation energy. 
We show that in order to have stable and analytic solutions, and for $f$ to 
satisfy the homogeneous scaling constraint, the power is restricted 
to $1/4 \leq \alpha \leq 3/4$. 
Applying a reasonable ansatz for the momentum distribution
and the lower bound for the exchange-correlation energy
tightens the physical regime further to $0.64 \lesssim \alpha \leq 0.75$.
\end{abstract}

 
\maketitle

\section{Introduction}

Reduced density-matrix functional theory (RDMFT) is
an appealing method to tackle the many-particle problem
through the one-body reduced density matrix (1-RDM),
which is obtained by integrating out all but two spatial 
coordinates in the many-body wave function.~\cite{lowdin,gilbert}
Computationally, RDMFT is significantly more demanding
than the more common density-functional theory~\cite{dft} (DFT) whose 
key quantity is the single-particle density, i.e., the
diagonal of the 1-RDM. On the other hand, RDMFT has been
shown to improve considerably from the present capabilities 
of DFT, especially when applied to strongly correlated
systems.~\cite{giesbertz}

As briefly reviewed below, the development of energy 
functionals of the 1-RDM is still at a relatively early
stage. In connection with this process, exact and general
constraints~\cite{levy}
of RDMFT have been studied in the case of the three-dimensional 
homogeneous electron gas (3DEG) which, as shown by Cioslowski
and Pernal,~\cite{kasia,kasia2} provides a natural framework
to obtain constraints to the 1-RDM functionals. More recently, 
the practical performance of different functionals for the 3DEG 
-- including
the physically appealing power functional~\cite{sharma} -- 
has been examined by Lathiotakis 
{\em et al.}~\cite{nek1,nek2}

Until now, developments of RDMFT in two dimensions (2D) have been
very scarce despite the large variety of strongly 
correlated 2D many-electron systems such as 
semiconductor quantum dots, quantum Hall devices,
and Aharonov-Bohm interferometers. In these systems
the movement of particles is essentially restricted to 
a plane, so that the quantum mechanical 
degrees of freedom have been frozen in the third
(off-plane) direction. Experience within 
DFT has shown that 2D functionals (where, e.g., the
density scaling is different from that in 3D) are needed 
to capture the physical properties of semiconductor
quantum dots.~\cite{rogge} In recent
years, many developments in DFT have been made in that 
direction.~\cite{stefano,constantin,constantin2} 
In RDMFT, however, the only work that addresses
2D systems have been done by Harju and T\"ol\"o,~\cite{harju}
who tested several 1-RDM (and also 2-RDM) functionals 
for quantum Hall droplets at high magnetic fields.

In this paper we aim at bridging the gap between the
methodological potential of RDMFT and the 
interest in 2D systems in the condensed matter community. 
In particular, we use the
homogeneous 2D electron gas (2DEG) as the playground
to obtain exact constraints to the 1-RDM functionals
applied in 2D. In the derivation we apply the procedure of 
Cioslowski and Pernal~\cite{kasia} mentioned above.
The existence of a stable solution and the 
homogeneous scaling constraint bring the first
density-independent restrictions to the scaling parameter.
Further density-dependent constraints are
introduced through the bounds on the natural orbitals
and on the exchange-correlation (xc) energy. Finally
we show that, apart from the possibility for border-minima solutions,
the M\"uller~\cite{muller} and Goedecker-Umrigar~\cite{gu}
functionals of the scaling $f(n,n')=(n n')^{1/2}$
are not valid for the 2DEG of any density,
and that the power functional with $f(n,n')=(n n')^{\alpha}$
is applicable only at $0.64 \lesssim \alpha \leq 0.75$.

\section{Reduced density-matrix functional theory}

In a system of $N$ interacting electrons the total energy 
of the ground state can be expressed as a sum of the kinetic, 
external, and electron-electron (e-e) interaction energy, i.e.,
\be
E_{\rm tot}[\Psi] = T[\Psi] + E_{\rm ext}[\Psi] + E_{\rm int}[\Psi],
\label{etot}
\ee
where the total energy is a functional of the $N$-electron 
ground-state wave function $\Psi$. According to Gilbert's theorem,
there is a one-to-one correspondence between
$\Psi$ and the one-body reduced density matrix (1-RDM),
\be
\gamma(\vr,\vr') = \Psi^*(\vr')\Psi(\vr) = \sum_{i=1}^\infty n_i \varphi^*_i(\vr')\varphi_i(\vr),
\ee
where $\varphi_i$ are natural orbitals and $n_i$ the corresponding
occupation numbers, which have values $0 \leq n_i \leq 1$ and 
sum up to $N$. In Eq.~(\ref{etot}), the two first terms have
simple expressions as functionals of $\gamma$, i.e.,
\be
T[\gamma] = \int d\vr \int d\vr' \delta(\vr-\vr')
\left[-\frac{1}{2}\nabla^2_\vr\right]\gamma(\vr,\vr')
\label{T}
\ee
and
\be
E_{\rm ext}[\gamma] = \int d\vr \int d\vr' \delta(\vr-\vr') V_{\rm ext}(\vr,\vr') \gamma(\vr,\vr').
\ee
In contrast, the e-e interaction term in Eq.~(\ref{etot}) has
a more complex expression,
\be
E_{\rm int}[\gamma] = \min_{\Psi\rightarrow\gamma}\frac{1}{2}\int d\vr \int d\vr' \frac{\rho_2(\vr,\vr')}{|\vr-\vr'|},
\label{int}
\ee
where the minimization is performed over all normalized and 
antisymmetric $\Psi$ that yield $\gamma$. Here the pair density is
defined by
\bea
\rho_2(\vr,\vr') & = & N(N-1)\int\ldots\int d\vr_3\ldots d\vr_N \nonumber \\
& \times & \Psi^*(\vr,\vr',\vr_3\ldots,\vr_N)\Psi(\vr,\vr',\vr_3\ldots,\vr_N),
\eea
which can be rarely calculated in practice.

Now, practical use of RDMFT requires {\em approximations} on 
$E_{\rm int}[\gamma]$ given in Eq.~(\ref{int}). At this stage, it is useful to partition the term as
\be
E_{\rm int}[\gamma] = E_{H}[\gamma] + E_{xc}[\gamma],
\label{comp}
\ee
where
\be
E_H[\gamma] = \frac{1}{2}\int d\vr \int d\vr' \frac{\gamma(\vr,\vr)\gamma(\vr',\vr')}{|\vr-\vr'|}
\ee
is the Hartree energy corresponding to the classical 
electrostatic energy calculated from the diagonal of
$\gamma$, i.e., the single-particle density $\rho(\vr)=\gamma(\vr,\vr)$.
The second term in Eq.~(\ref{comp}) is the xc
energy. It is important to note that this is {\em not} the same as the xc 
energy in the conventional DFT. The difference is due to the exact expression of the kinetic energy 
in RDMFT [Eq.~(\ref{T})] in contrast with the approximate DFT expression 
employing the Kohn-Sham orbitals.

The simplest approximation for $E_{xc}$ is given by the Hartree-Fock (HF) term
\bea
E_{xc}^{\rm HF}[\gamma] & = & -\frac{1}{2}\int d\vr \int d\vr' \frac{\gamma(\vr,\vr')\gamma(\vr'\vr)}{|\vr-\vr'|} \nonumber \\
                      & = & -\frac{1}{2}\sum_{i,j}^\infty  \int d\vr \int d\vr' \nonumber \\
                      & \times & n_i n_j \frac{\varphi^*_i(\vr)\varphi^*_j(\vr')\varphi_j(\vr)\varphi_i(\vr')}{|\vr-\vr'|},
\eea
which corresponds to the exchange energy of RDMFT. Most further approximations
in RDMFT have emerged from this expression with an aim to account for the
correlation energy which is omitted in the HF term. These
functionals can be expressed in a form
\bea\label{exc}
E_{xc}[\gamma] &   =    &  -\frac{1}{2}\sum_{i,j}^\infty  \int d\vr \int d\vr' \nonumber \\
              & \times &  f(n_i,n_j) \frac{\varphi^*_i(\vr)\varphi^*_j(\vr')\varphi_j(\vr)\varphi_i(\vr')}{|\vr-\vr'|},
\eea
where the difference from the HF expression is embedded in the function 
$f(n_i,n_j)$. 

Various approximations for $f(n_i,n_j)$ have been introduced, many of
them based on a form $f(n_i,n_j)=(n_i n_j)^\alpha$, where $\alpha=1$
corresponds to the HF case. One of the first approximations is the
M\"uller functional with $\alpha=1/2$ (Ref.~\onlinecite{muller}).
The factor is the same in the
Goedecker-Umrigar functional~\cite{gu} that removes 
the orbital self-interaction. 
Further hierarchical corrections, where $f(n_i,n_j)$ depends
on the orbital occupancies, have been introduced by Gritsenko 
{\em et al.}~\cite{bbc} They are commonly known as BBC1, BBC2, and BBC3
functionals according to corrections to the Buijse-Baerends 
functional.~\cite{buijse} 

More recently, the so-called power functional was introduced by 
Sharma {\em et al.}~\cite{sharma} In that functional 
$\alpha$ is introduced as a 
mixing parameter between the HF and M\"uller approximations, 
i.e, $1/2\leq\alpha\leq 1$. It was found that the optimal $\alpha$ varies 
between $0.525$ (stretched $H_2$) and $0.65$ (solids), and the best overall 
fit for the 3D homogeneous electron gas was found at 
$\alpha=0.55\ldots 0.58$ (Ref.~\onlinecite{nek2}).
In 2D, RDMFT has been applied to quantum Hall droplets by T\"ol\"o and
Harju~\cite{harju} who found that in in many cases the power functional 
yields reasonable results with $\alpha\sim 0.65\ldots 0.7$, but also
clear limitations of the functional were found.

\section{Two-dimensional homogeneous electron gas}\label{2deg}

In 2DEG we can consider a positive background charge 
compensating for the electrostatic (Hartree) energy,
so that the total energy [Eq.~(\ref{etot})] consists
of the kinetic and xc components alone, i.e.,
\be
E^{\rm 2DEG}_{\rm tot}[\gamma] = T[\gamma] + E_{xc}[\gamma],
\label{etot2deg}
\ee
where
\be
T[\gamma] = -\frac{1}{2}\sum_{\sigma=\uparrow,\downarrow}\sum_p n_\sigma(\vk_p)\int d\vr \,\varphi^*_{p\sigma}(\vr) \nabla^2 \varphi_{p\sigma}(\vr)
\label{T2}
\ee
and, similar to Eq.~(\ref{exc}),
\bea\label{exc2}
E_{xc}[\gamma]  &    =   &  -\frac{1}{2}\sum_{\sigma=\uparrow,\downarrow}\sum_{p,q}^\infty  \int d\vr \int d\vr' f\left(n_\sigma(\vk_p),n_\sigma(\vk_q)\right) \nonumber \\
       & \times & \frac{\varphi^*_{p\sigma}(\vr)\varphi^*_{q\sigma}(\vr')\varphi_{q\sigma}(\vr)\varphi_{p\sigma}(\vr')}{|\vr-\vr'|},
\eea
where we have introduced $\vk_p$ as the wave vector of the $p$th 
spin-dependent natural orbital. In the 2DEG it can be now expressed 
as a plane wave,
\be
\varphi_{p\sigma}(\vr) = A^{-1/2}\exp(i\vk_p\cdot\vr)\chi_p(\sigma),
\ee
where $A$ is the area and $\chi$ is the spin component.
Now, it is straightforward to calculate the 
integrals in Eqs.~(\ref{T2}) and (\ref{exc2}) yielding, respectively,
$-|\vk_p|^2$ and $2\pi A^{-1}|\vk_p-\vk_q|^{-1}$. The summation
over plane waves can be replaced by an integration,
\be
\sum_p \rightarrow (4\pi^2)^{-1} A \int d\vk.
\ee
As a result of these replacements, 
we can express the total energy of the 2DEG
as a functional of the momentum distribution,
\bea
E^{\rm 2DEG}_{\rm tot}[n_\uparrow,n_\downarrow] & = & \frac{A}{8\pi^2}\sum_{\sigma=\uparrow,\downarrow}\int d\vk\,n_\sigma(\vk)|\vk|^2 \\
                       & - & \frac{A}{16\pi^3}\sum_{\sigma=\uparrow,\downarrow}\int d\vk \int d\vk' \, \frac{f\left(n_\sigma(\vk),n_\sigma(\vk')\right)}{|\vk-\vk'|}. \nonumber
\eea
The normalization constraint is given by an integral
\be
N_\sigma = \frac{A}{4\pi^2}\int d\vk \, n_\sigma(\vk) = A\,\rho_\sigma,
\label{N}
\ee
where $\rho_\sigma$ is the spin density (per area). We can 
define the areal energy density per spin simply by 
\be
\epsilon_\sigma[n_\sigma]=\frac{E^{\rm 2DEG}_{\rm tot}[n_\sigma]}{A} = t_\sigma[n_\sigma] + \epsilon^{\sigma}_{xc}[n_\sigma],
\label{energy1}
\ee
where $t_\sigma$ and $\epsilon^{\sigma}_{xc}$ are the kinetic and xc energy 
densities per spin, respectively. 

For brevity, we omit the spin index 
$\sigma$ in the following. Thus, it should be noted that in the rest 
of the paper 
$n$, $\rho$, $\epsilon$, $t$, and $\epsilon_{xc}$ refer to
quantities per spin-particle (with spin $\sigma$).
With this notation, particular care is needed for the formula 
of the lower bound on the xc energy
[Eq.~(\ref{bound}) below], which
is conventionally considered for the spin-unpolarized 2DEG with 
$\rho_{\rm tot}=\rho_\uparrow+\rho_\downarrow=2\rho_\uparrow$.

We can now seek for a stationary value for the functional
$\epsilon[n]$ with respect to variations in $n$, so that
the normalization in Eq.~(\ref{N}) is satisfied. The
normalization constraint introduces a Lagrange multiplier
$\mu$ in the variational (Euler-Lagrange) equation written as
\be
\frac{d\epsilon}{d n} - \mu\frac{d \rho}{d n} = 0,
\label{euler}
\ee
leading to
\be
\frac{1}{2}|\vk|^2 - \frac{1}{2\pi}\int d\vk' \, \frac{\frac{\partial}{\partial n(\vk)}f\left(n(\vk),n(\vk')\right)}{|\vk-\vk'|} = \mu.
\label{n}
\ee
It is important to note that in this paper we concentrate on 
{\em analytic} solutions of Eq.~(\ref{euler}).
Thus, we exclude the possibility for partial or full border minima,
where $n(\vk)$ is equal to 0 or 1 -- as allowed by the
$N$-representability condition -- but Eq.~(\ref{euler}) is not 
satisfied.~\cite{nek1}

\subsection{General constraints}\label{general}

Next we examine a set of constraints for the solutions of 
Eqs.~(\ref{energy1})-(\ref{n}). First, 
the (ground-state) solutions have to be stable
upon normalization. Hence, for a uniform scaling 
with a constant, positive $\lambda\neq 1$
we require
\be
\epsilon[\lambda^2 n(\lambda\vk)]>\epsilon[n(\vk)],
\ee
so that the first derivative with respect to $\lambda$ at $\lambda=1$
is zero,
\be\label{1stDer}
\frac{\partial}{\partial\lambda}\epsilon[\lambda^2 n(\lambda\vk)]\big|_{\lambda=1}=0,
\ee
and the second derivative is positive,
\be\label{2ndDer}
\frac{\partial^2}{\partial\lambda^2}\epsilon[\lambda^2 n(\lambda\vk)]\big|_{\lambda=1}>0.
\ee
As the second constraint, the function 
$f\left(n(\vk),n(\vk')\right)$ has to satisfy the 
homogeneous scaling,~\cite{levy}
\be\label{scaling}
f\left(\lambda^{2} n(\vk),\lambda^{2} n(\vk')\right) = \lambda^{2\beta} f\left(n(\vk),n(\vk')\right),
\ee
for all $n(\vk)$ and $n(\vk')$.
The kinetic-energy density scales as
$t\left(\lambda^{2}\,n(\lambda\vk)\right)=\lambda^{-2}\,t\left(n(\vk)\right)$, 
and, from Eq.~(\ref{scaling}) we find that 
the xc energy scales as 
$\epsilon_{xc}(\lambda^{2}\,n(\lambda\vk))=\lambda^{2\beta-3} \epsilon_{xc}(n(\vk))$.
Now, Eq. (\ref{1stDer}) leads to 
\be
t(\rho)=\frac{2\beta-3}{2\beta-1}\epsilon(\rho),
\label{tcond}
\ee
and with Eq.~(\ref{energy1}) to
\be
\epsilon_{xc}(\rho)=\frac{2}{2\beta-1}\,\epsilon(\rho).
\label{xccond}
\ee
By definition, the kinetic-energy density must be
positive, $t>0$, and the xc energy must be non-positive, 
$\epsilon_{xc}\leq 0$. It is then straightforward to show
with Eqs.~(\ref{2ndDer}), (\ref{tcond}), and (\ref{xccond})
that both these conditions are satisfied when
 \be
\frac{1}{2} < \beta < \frac{3}{2}.
\label{betacond}
\ee
The range is slightly larger than in 3D,
where the corresponding result is
$2/3 < \beta_{\rm 3D} < 4/3$ (Ref.~\onlinecite{kasia}).

From Eq.~(\ref{scaling}) we may immediately conclude 
that in the case of a power functional, 
$f(n_i,n_j)=(n_i n_j)^\alpha$, the power $\alpha$ is 
restricted to 
\be
\frac{1}{4} < \alpha < \frac{3}{4}.
\ee
Hence, e.g., the M\"uller functional with $\alpha=1/2$
(see the next section) satisfies the stability and 
homogeneous scaling constraints. 
It is important to note, however, that there
are further restrictions for the density $\rho$, which in
fact appear to be prohibitive to the M\"uller functional.
These limitations arise from (i) how the specific form of
the momentum distribution
satisfies the condition $0\leq n(\vk) \leq 1$,
and from (ii) the universal lower bound for $\epsilon_{xc}$.
The latter condition corresponds to
the 2D counterpart of the Lieb-Oxford bound~\cite{lo} 
whose existence was rigorously proven 
by Lieb, Solovej, and Yngvason.~\cite{lsy} 
Recently, however, the {\em tightest} 
form for this bound was suggested 
through nonrigorous but physical arguments.~\cite{ourbound}
The bound was found to correspond to the 
low-density limit of the 2DEG -- the same system as in 
consideration here -- and  for the spin-unpolarized 2DEG
it can be written as
\be
2\,\epsilon_{xc}\geq -C\,(2\rho)^{3/2},
\label{bound}
\ee
where $C=1.96$ (Ref.~\onlinecite{ourbound}).
The factor of two on both sides of Eq.~(\ref{bound})
results from the per-spin notation (see above).

\subsection{Limits of the M\"uller functional}

We examine the constraints of the M\"uller functional,
\be
f\left(n(\vk),n(\vk')\right) = \left(n(\vk) n(\vk')\right)^{1/2},
\ee
which in fact coincides with the Goedecker-Umrigar
functional in the case of the 2DEG, since the
self-interaction terms vanish for plane-wave
orbitals. Now we have $\beta=1$ and use an ansatz
\be
n(\vk) = \rho\,\eta(\vk),
\label{mulleransatz}
\ee
where $\eta$ is independent of the density and satisfies the 
normalization 
\be
\int d\vk\,\eta(\vk)=4\pi^2.
\label{etanormalization}
\ee

We can solve the variational equation 
(\ref{n}) that becomes,
\be
\frac{1}{2}|\vk|^2\eta(\vk)^{1/2} - \frac{1}{4\pi}\int d\vk' \, \frac{\eta(\vk')^{1/2}}{|\vk-\vk'|}=\mu\eta(\vk)^{1/2}.
\label{variationalmuller}
\ee
Substituting  $\eta(\vk)=\phi(\vk)^{2}$ and taking the Fourier transform yields
\be
-\frac{1}{2}\nabla^2 {\tilde \phi}(\vr)-\frac{1}{2}\,{\tilde \phi}(\vr)\frac{1}{|\vr|}=\mu \, {\tilde \phi}(\vr),
\ee
which has a simple solution,
\be
{\tilde \phi}(\vr) = \sqrt{8\pi}\,e^{-|\vr|}
\ee
with $\mu=-1/2$. Taking an inverse Fourier transform leads to
\be
\eta(\vk)=8\pi(1+|\vk|^2)^{-3},
\label{mullermomentum}
\ee
and finally, according to Eq.~(\ref{mulleransatz}) 
the momentum distribution of the M\"uller functional
in the 2DEG reads
\be
n(\vk)=8\pi\rho(1+|\vk|^2)^{-3}.
\label{nmuller}
\ee

Now we go back to constraints (i) and (ii) mentioned at the end
of the previous section.
First, from the requirement $0\leq n(\vk)\leq1$ it follows that
$\rho\leq (8\pi)^{-1}$ for the spin-density. In terms of
the commonly used density parameter $r_s$ (Wigner-Seitz radius)
for the spin-unpolarized 2DEG the condition reads
\be
r_s\equiv (2\pi\rho)^{-1/2}\geq 2.
\label{rs1}
\ee
To obtain the second constraint (ii) we first
employ Eqs.~(\ref{tcond}) and (\ref{xccond}) and find
$t=\rho/2$, $\epsilon_{xc}=-\rho$,  and $\epsilon=-\rho/2$.
Thus, again for the spin-unpolarized 2DEG, 
we get from Eq.~(\ref{bound}) another condition 
\be
r_s \leq \frac{C}{\sqrt{\pi}} \approx 1.1.
\label{rs2}
\ee
We can immediately see that conditions ({\ref{rs1}}) and 
({\ref{rs2}}) are exclusive. Thus, the 
M\"uller functional is not valid for the 2DEG
of any density, apart from the possibility of 
border-minima solutions (see the end of Sec.~\ref{2deg}).

\subsection{Limits of the power functional}

Finally we examine the limits of the power functional
\be
f\left(n(\vk),n(\vk')\right) = \left(n(\vk) n(\vk')\right)^{\beta/2} = \left(n(\vk) n(\vk')\right)^{\alpha}.
\ee
In contrast with the M\"uller functional with $\beta=1$ 
we cannot analytically solve the variational equation (\ref{n})
in order to obtain the exact momentum distribution. 
However, we may try to find a reasonable ansatz for the 
momentum distribution. First, we examine how the Euler-Lagrange
equation (\ref{euler}) scales with the density and 
separate the density-dependent part from $n(\vk)$ as
\be
n(\vk)= \rho^{1/(2\beta-1)}\eta\left(\rho^{\frac{1-\beta}{2\beta-1}}\vk\right),
\label{ansatz}
\ee
where $\eta$ is independent of the density. Following
the strategy of Cioslowski and Pernal,~\cite{kasia} 
we choose a parametrized {\em trial} function for $\eta$
similar to that of the M\"uller functional [Eq.~(\ref{mullermomentum})],
\be
{\bar \eta}(\vk) = D(1+\zeta |\vk|^2)^{-3/\beta},
\label{momentumpower}
\ee
where $D=4\pi\zeta(3\beta^{-1}-1)$ is the normalization
constant obtained from Eq.~(\ref{etanormalization}).
The parameter $\zeta$ is now solved such that the
total energy density, which has a form
\be
\epsilon(\rho) = I_\epsilon \, \rho^{\frac{2\beta-2}{2\beta-1}+1}
\label{energy2}
\ee
is minimized. As pointed out in Ref.~\onlinecite{kasia}, this
forms an {\em upper bound} to the integral
\bea
\label{iepsilon}
 I_\epsilon & = & \frac{1}{8\pi^2}\int d\vk\,\eta(\vk)|\vk|^2 \nonumber \\
                       & - & \frac{1}{16\pi^3}\int d\vk \int d\vk' \, \frac{f\left(\eta(\vk),\eta(\vk')\right)}{|\vk-\vk'|},
\eea
which is independent of the density, and
has the absolute, stable minimum for the 
unkown {\em exact} momentum distribution. 
Thus, we have an inequality $I_\epsilon\leq {\bar I}_\epsilon$, where 
\bea\label{bari}
{\bar I}_\epsilon & = & \min_\zeta\Bigg\{\frac{\beta}{2\zeta(3-2\beta)} - 4^{\beta-2}\pi^{\beta-3}\left(\frac{3\zeta}{\beta}-\zeta\right)^\beta F(\zeta)\Bigg\} \nonumber 
\eea
with an integral
\be
F(\zeta) = \int d\vk \int d\vk' \, \frac{(1+\zeta|\vk|^2)^{-3/2}(1+\zeta|\vk'|^2)^{-3/2}}{|\vk-\vk'|}.
\ee
The integral is similar to the Hartree energy integral with a ``density'' distribution
$(1+\zeta|\vk|^2)^{-3/2}$. In the lack of an analytic solution (although we do not preclude
its possible existence), we solve $F(\zeta)$ numerically by taking a Fourier transform 
and using the convolution theorem implemented in the {\tt octopus} code.~\cite{octopus}
We find $F(\zeta) = \gamma\,\zeta^{-3/2}$ with $\gamma\approx 19.74$. 
After rewriting Eq.~(\ref{bari}) and differentiating with respect to $\zeta$
we obtain the minimum at
\be
\zeta_m(\beta) = \left[\frac{a}{b(\beta-3/2)}\right]^{\frac{1}{\beta-1/2}}
\ee
with $a(\beta)=(6\beta^{-1}-4)^{-1}$ and $b(\beta)=4^{\beta-2}\pi^{\beta-3}(3\beta^{-1}-1)^\beta \gamma$.
Thus, ${\bar I}_\epsilon=a\,\zeta_m^{-1}-b\,\zeta_m^{-3/2}$ is obtained by inserting $\zeta_m$ to Eq.~(\ref{bari}).

Let us now consider the constraints (i) and (ii) for the
allowed values of $n(\vk)$ and $\epsilon_{xc}$, 
respectively (see the end of Sec.~\ref{general}).
First, from Eqs.~(\ref{ansatz}) and (\ref{momentumpower}) we obtain
\be
0 \leq {\bar n}(\vk)=\rho^{1/(2\beta-1)} D (1+\zeta_m |\vk|^2)^{-3/\beta} \leq 1
\ee
leading to
\be
\rho \leq D^{1-2\beta} = \left[4\pi\zeta_m (3\beta^{-1}-1)\right]^{1-2\beta}.
\label{rho1}
\ee
Secondly, from Eqs.~(\ref{energy2}), (\ref{xccond}), and (\ref{bound}) we obtain
\be
\frac{4 {\bar I}_\epsilon}{2\beta-1}\,\rho^{\frac{2\beta-2}{2\beta-1}+1} 
\geq -C (2\rho)^{3/2},
\ee
yielding another condition for the density,
\be
\rho\geq \left[2^{-1}C^2 {\bar I}_\epsilon^{-2}(2\beta-1)^2\right]^{\frac{2\beta-1}{2\beta-3}}.
\label{rho2}
\ee
Combining Eqs.~(\ref{rho1}) and (\ref{rho2}) leads to a single 
constraint for $\beta$ which has a simple form,
\be
\beta \geq \frac{3}{2\pi\left(C/\gamma\right)^{2/3}+1} \approx 1.28.
\ee

\begin{figure}
\includegraphics[width=0.85\columnwidth]{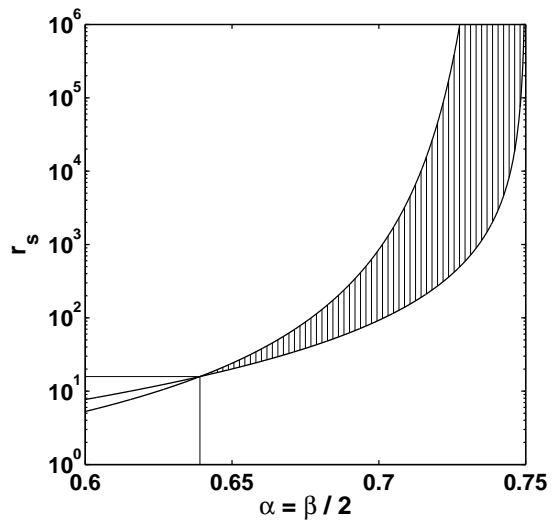}
\caption{Allowed densities (shaded region) of the
two-dimensional electron gas as a function of
 $\alpha=\beta/2$, which is physically limited to  
$0.64 \lesssim \alpha \leq 0.75$.
The limiting curves are given by
Eqs.~(\ref{rho1}) and (\ref{rho2}). The maximum possible 
density (minimum Wigner-Seitz radius $r_s$) having
an admissable power ($\alpha\approx 0.64$) is 
$r_s\approx 15$. 
}
\label{fig1}
\end{figure}

Now, keeping in mind the physical limits of $\beta$
given in Eq.~(\ref{betacond}),
we get a condition $1.28\lesssim\beta\leq 1.5$.
Correspondingly, the power functional
satisfies the stability (in terms of the analyticity of the minimum) 
and homogeneous scaling constraints 
{\em and} constraints (i) and (ii), when the power is restricted to
\be
0.64 \lesssim \alpha \leq 0.75.
\label{criterion}
\ee
It should be noted, however, that the lower
limit, for example, is valid only for a single density,
i.e., the one that satisfies the {\em equality} conditions in 
both Eqs.~(\ref{rho1}) and (\ref{rho2}).
At larger $\alpha$, the a range for allowed densities 
increases as visualized in Fig.~\ref{fig1}.
The maximum possible density, for which an
admissable power exists, corresponds to 
a minimum value $r_s\approx 15$ for the density parameter.
This density range is too low for the most 2D applications in, e.g.,
quantum dot and quantum Hall physics dealing with
desities in the range $0.1 \lesssim r_s \lesssim 10$.

Despite the strict conditions for $\alpha$ and the
corresponding densities, it 
is interesting to note that the optimal power
for 2D quantum Hall droplets was found to be 
$0.65\ldots 0.7$ (Ref.~\onlinecite{harju}), which
coincides with the allowed range obtained here. 
However, it is important to bear in mind that our
analysis for the 2DEG does not include (i) the possibility for 
border minima, and (ii) the evaluation of the {\em accuracy} of the power 
functional in comparison with the exact xc energy of 
the 2DEG known through quantum Monte 
Carlo calculations.~\cite{tanatar,attaccalite}
These issues will be addressed in future works.
We also point out that the validity of our ansatz
momentum distribution in Eq.~(\ref{ansatz})
could be further evaluated.

\section{Summary}

In summary, we have examined the constraints
of reduced density-matrix functionals in the 
description of the homogeneous two-dimensional 
electron gas, which is the base for a large spectrum
of applications in low-dimensional physics, e.g.,
in the quantum Hall regime. As our main result,
we have found that the power of the scaling function 
$f(n,n')=(n n')^\alpha$ is physically limited to
$0.64 \lesssim \alpha \leq 0.75$. The result
has been preceded by a thorough analysis of
how $\alpha$ is restricted and affected by
(i) the existence of stable solutions with analytic 
minima, (ii) 
the homogeneous scaling constraint for $f(n,n')$,
(iii) the allowed values for $n(\vk)$,  and (iv) 
the lower bound of the exchange-correlation energy.
Yet another issue to
be addressed in the future is the possibility
for border-minima solutions as well as the
the practical accuracy
of the power functional in comparison with
exact results when applied --
within the constraints addressed here --
to different systems, first and foremost
to the two-dimensional electron gas of
different densities.
In general, we hope that our analysis
serves as a useful guideline in the development
of density-matrix functionals in two dimensions.

\begin{acknowledgments}
We thank Katarzyna Pernal, Klaas Giesbertz, Robert van 
Leeuwen, and Florian Eich for helpful comments and 
useful discussions. This work has been supported by
the Academy of Finland.
\end{acknowledgments}


\end{document}